\documentclass[structabstract]{aa}
\usepackage{graphicx}
\usepackage{txfonts}
\begin{document}
   \title{A new method of determining distances to dark globules}

   \subtitle{The distance to  B\,335}

   \author{S. Olofsson
          \inst{1}
          \and
          G. Olofsson\inst{1}\fnmsep\thanks{Based on observations collected
          at the European Southern Observatory, Chile (ESO programme 077.C-0524(A))}
       }

   \institute{Stockholm Observatory, Stockholm University, Astronomy Department
              AlbaNova Research Center, SE-106 91 Stockholm\
              \email{sven@astro.su.se}
        % \and
           %  University of Alexandria, Department of Geography, ...\\
            % \email{c.ptolemy@hipparch.uheaven.space}
             %\thanks{The university of heaven temporarily does not
            %         accept e-mails}
             }

   \date{Received  23 December 2008; accepted 13 February 2009}

% \abstract{}{}{}{}{}
% 5 {} token are mandatory

  \abstract
   {The distance to an isolated dark globule is often unknown and yet crucial for
   understanding its properties, in particular its mass. A new approach to this problem is discussed}
   {The purpose of the present paper is to investigate how well the distances of
   more or less reddened field stars can be determined by using multi-colour imaging.}
   {We observed a test globule, B\,335 in U, B, g, r, and I, and together
   with the 2MASS survey, this data set gives a well-defined spectral energy distribution (SED) of a large number of stars.
   The SED of each star depends on the interstellar extinction, the distance to the star, and its intrinsic SED.
   As we had good reasons to suspect that the wavelength dependence of the extinction (the reddening) changes from
   the outskirts of the globule to the central parts, we did not assume any specific reddening law.
   Instead, we use a scheme that allows independent determination of the extinction in each line of sight as
   determined by groups of adjacent stars. The method is based on the use of
   stellar atmospheric models to represent the intrinsic SEDs of the stars.
%  We use synthetic colours to represent the intrinsic SEDs of the stars.
   Formally, it is then possible to determine the spectral class of each star and thereby its distance. For some of the stars we
   have optical spectra,  allowing us to compare the photometric classification to the spectrometric. }
   {As expected, the main problem is that there are few stars found within each
   distance bin for the small field size defining a typical dark globule. However, the characterisation of
   the extinction and photometric classification give consistent results and we can identify one star at the front side of the globule.
   It has a photometric distance of 90\,pc. The closest star behind the B\,335 globule has a
   distance of only $\approx 120$\,pc and we therefore determine the distance to B\,335 as 90--120\,pc.  Our deep U image shows
   a relatively bright south-western rim of the globule, and we investigate whether it might be due to a local enhancement of the radiation field.
   A candidate source, located 1.5 arcminutes outside our field, would be the field star, HD\,184982. This star has
   an entry in the Hipparcos Catalogue and its distance is 140--200\,pc. However, we come to the conclusion that
   the bright SW rim is more likely due to the wing of the point-spread-function (PSF) of this star.}
   {}


   \maketitle
%
%________________________________________________________________

\section{Introduction}

It is in principle easy to determine the distance to a dark cloud: it is just a matter of finding
 a star  just in front of the cloud and another  closely behind it. Or, even better, a star close
enough to the cloud to give rise to a reflection nebula. The closest cloud complexes, like that in
Taurus/Aurigae, cover large regions on the sky and thanks to the Hipparcos mission (\cite{Hipparcos}) there are
a number of stars with known distances within the field. The remaining task is that of placing the stars
in front of or behind the cloud. The traditional way is to look for the onset of
interstellar reddening (see Lombardi et al. \cite{Lombardi} for a recent example). As an
attractive alternative, the onset of polarisation can be determined (Alves \& Franco \cite{Alves}).
For small isolated clouds, however, there are too few (if any)  Hipparcos stars within the field,
and the distances to the stars must be determined by spectroscopic/photometric means.  The reason
why it is important to determine the distances to such clouds, including dark globules, is their
role of hosting isolated star formation. Thus,  important properties as the luminosity of the
forming star and the mass and the radius of the cloud cannot be determined without knowing the distance.

Maheswar \& Bhatt (\cite{Maheswar}) have estimated the distances to nine dark globules using
V, R and I photometry in combination with the J, H and K photometry from the 2MASS catalogue
 (Skrutskie et al. \cite{2MASS}). By assuming a standard interstellar extinction law (valid
 for \emph {diffuse} clouds) and adjusting visual extinction, A$_V$, until the SED fits that
 of a \emph{main sequence} (MS) star, these authors determine the extinction and the distance
 for each star. There are two uncertain assumptions in their approach, which may have strong
 impact on the estimates of the distances. The first is the assumption of a standard extinction
 law, while it is known that the $R_{V}$ value ($R_{V}$=A$_{V}$/E(B-V)) changes from typically
 $R_{V}$=3.1 in the diffuse ISM to $R_{V}\approx 5$ in molecular clouds
 (see e g Ossenkopf \& Henning \cite{Ossenkopf}, Strafella et al \cite{Strafella}). The second
 is the assumption that all the stars are MS stars. In the present paper we discuss these
 problems and present a more elaborate method which is based on imaging in eight filters.

As a test case we use the well studied dark globule B\,335. It hosts a protostar heavily
obscured by a flattened core structure, including an accretion disk seen edge-on
(G\aa lfalk \& Olofsson \cite{Gaalfalk}) . In most of the numerous investigations
of this source, a distance of 250\,pc as given by Tomita et al. (\cite{Tomita})
is assumed.  Stutz et al. (\cite{Stutz}) have recently revised the distance estimate
by using archive photometry of stars within a radius of 20\arcmin. Their revised
distance is 150\,pc (60--200\,pc). The diameter of B\,335 as seen in the optical
is just $\approx$ 4\arcmin\,. CO observations (Frerking  et al. \cite{Frerking}), however,
show that it extends far to the NE, including stars within such a large radius as 20\arcmin.
%__________________________________________________________________

% 1112093 19 38 24   0 31  5   0.637  0.283  0.370  0.397  0.766    0.629  0.283  0.405  0.384  0.761   0.22   0.21  1.483e-003  6212.0   F7V
\begin{table}
\begin{minipage}[t]{\columnwidth}
\caption{Observation log}
\label{obslog}
\renewcommand{\footnoterule}{}  % to avoid a line before footnotes
\begin{tabular}{llllll}
\hline \hline
object &camera &filter &center &exp time&exp time\\
& & &  wavel   &per frame &total\\
& & &$ [nm]$&[s]&[s]\\
B\,335 &EMMI-blue &U602  &353.96 &720 &26600\\
B\,335 &EMMI-red  &Bb605  &413.22 &720 &7900\\
B\,335 &EMMI-red  &g772  &508.86 &480 &4700\\
B\,335 &EMMI-red  &r773  &673.29 &300 &2100\\
B\,335 &EMMI-red  &I610  &798.50 &300 &2400\\
B\,335 &NOTCAM\footnote{co-added image provided by M. G{\aa}lfalk\\} &Ks &2144. &300  &2800\\
B\,335ref &EMMI-blue &U602 &353.96 &720 &2200\\
B\,335ref &EMMI-red &Bb605 &413.22 &720 &800\\
B\,335ref&EMMI-red &g772 &508.86 &480 &500\\
B\,335ref &EMMI-red &r773 &673.29 &300 &300\\
B\,335ref &EMMI-red &I610 &798.50 &300 &300\\
\hline\\
\end{tabular}
\end{minipage}
\end{table}

\section{Observations and data reduction}
B\,335 (RA(2000) = 19 37 0.8, DEC(2000) = 7 34 11 ]) and a reference field (centered at 19 34 10., 07 37 05)
were observed using the ESO/NTT at La Silla during four
nights 2006-06-27--30. The reference field was chosen to have the
same galactic latitude as  B\,335. The observational details are given in Table~\ref{obslog}.
The EMMI images have been complemented by a K$_s$ image taken with the 2.5\,m
Nordic Telescope on La Palma (total exposing 48 min; cf. G\aa lfalk et al. \cite{Gaalfalk}).\\
Furthermore, spectra from ten of the stars in the field were measured
with NTT-EMMI in RILD-mode with a resolution of $\thicksim 600$
in the wavelength range 550 - 1000 nm (exposure times between 300 and 1000 s,
slit 1.5\arcsec\,and 5 \arcsec).\\
All object exposures were preceded and followed by standard star (Landholt equatorial
standards SA111-2093) observations in respective filter. The basic reductions (bias subtraction,
dark correction, flat-fielding, cosmic ray reduction) were carried
out using standard IRAF routines. Then the IRAF astrometric programs
were used for registering and co-adding the images using Skyview (http://skyview.gsfc.nasa.gov)
facilities for astrometric data on stars common to all images of an
object. The summation of the frames were weighted by the
inverse of the errors of the stars in the middle magnitude range.
The star finding program 'sextractor' were then used to tabulate
stars and positions. Photometry on all co-added images were then
performed using the DAOPHOT photometry package.\\
In addition,  our observations were combined with data from the 2MASS Point
Source Catalog (PSC) (Skrutskie et al. \cite{2MASS}) thus extending the
stellar data with the J-H-K measurements (where these exist) in the
B\,335 and the reference field.
Furthermore, object field data in wavelengths around $3.6, 4.5, 5.8, 8.0 \mu$
from the Spitzer data archive were used. The Spitzer archived images were
analyzed with the Mopex-software package
(see Makovoz \& Khan  \cite{Makovoz-a}, Makovoz \& Marleau  \cite{Makovoz-b}, and Masci et al \cite{Masci})
to give the stellar magnitudes.
 Finally, tables for B\,335 and the reference field containing object positions,
magnitudes and magnitude errors in all filters were compiled. The field of
view differs between the red and the blue arm of EMMI and  the 4\arcmin field of
NOTCam was not exactly centered at the same position as the EMMI field. This results
in the observation field which is slightly less than 6\,arc-minutes. \\

 \section{Results and interpretation}

  \subsection{Setting the scene}

\begin{figure}
   \centering
   \includegraphics[width=9cm]{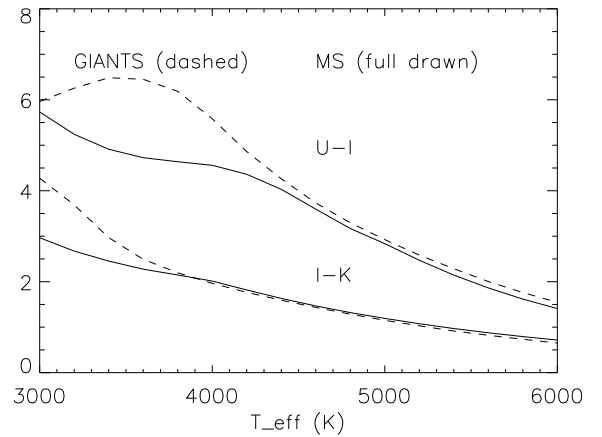}
      \caption{Unreddened U-I and I-K colours for giants (full lines) and main
        sequence stars (dashed) as a function of effective temperature based
        on the model atmospheres of Hauschildt et al.(1999). K and M giants can be
        separated from main sequence stars with the same temperature
       by measuring the U - I colour index. The I - K colour index is
       sensitive to gravity for late M stars.}
         \label{aa11574fig01}
   \end{figure}

\begin{figure}
   \centering
   \includegraphics[width=9cm]{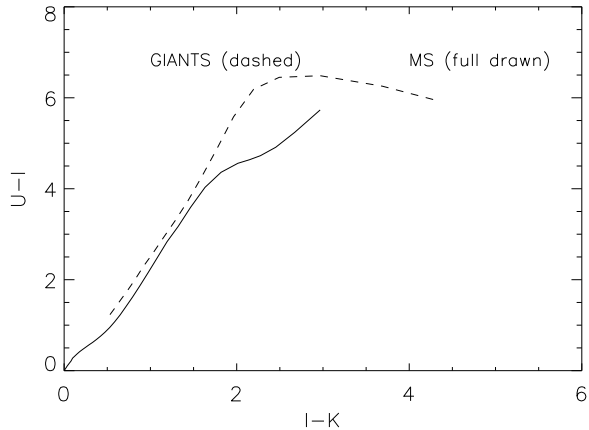}
      \caption{ The (U-I) colour versus (I-K) colour plot has the potential
      to separate K and M giants from main sequence
      stars - provided the interstellar reddening vector E(U-I)/E(I-K)  is
      less than the slope of the straight part of the curves ($\approx 2.7$).
              }
         \label{aa11574fig02}
   \end{figure}

Our observations basically gave the SEDs for a large number of stars, more or less hidden by the globule.
Our purpose was to identify stars immediately behind and in front of  the globule - if there are any.
To do so, we needed the intrinsic colours of the different stars and their corresponding absolute magnitudes.
The single most important parameter that determines the colour of a star is its temperature, but
for a given temperature the luminosity is very different for a dwarf as compared to a giant or supergiant star.
supergiants are so rare that the chance of finding one within such a small field, a few degrees away from
the Galactic plane, is small. As will be discussed below, giant stars cannot be excluded on the same grounds, and we
had to look for means to separate giants from dwarfs. To avoid the difficulties to interpolate between different
photometric systems, we chose to use synthetic colours to represent the intrinsic colours of the stars. We chose
the {\it Next Generation} model atmospheres provided by Hauschildt et al. \cite{Hauschildt}. In order to keep the parameters
within reasonable limits, we restricted the range of surface gravity to one value typical for MS stars
(log\,g = 4.5) and one typical for giants (log\,g = 0). We also restricted the metal
abundance to solar (typical for Population\,I, which dominates in most directions along the Galactic plane). As one
may suspect, it is hard to separate dwarfs from giants in most combinations of colour indices. However, there
is a tendency for the cool (K and M) giants to be redder than the dwarfs for the same T$_{eff}$ and this effect is
pronounced in the  U--I colour index, but also seen in eg. the I--K index for late M stars (see Fig~\ref{aa11574fig01}).
By combining these two indices in a colour--colour plot, we should be able to separate K and M giants from dwarfs (Fig~\ref{aa11574fig02}).

%%AA/2008/11574
%                                                One column figure
%______________________________________________________________
%                                                              fig 1

%
%______________________________________________________________
%                                                              fig 2

%
%______________________________________________________________
%                                                              fig 3

   \begin{figure}
   \centering
   \includegraphics[width=9cm]{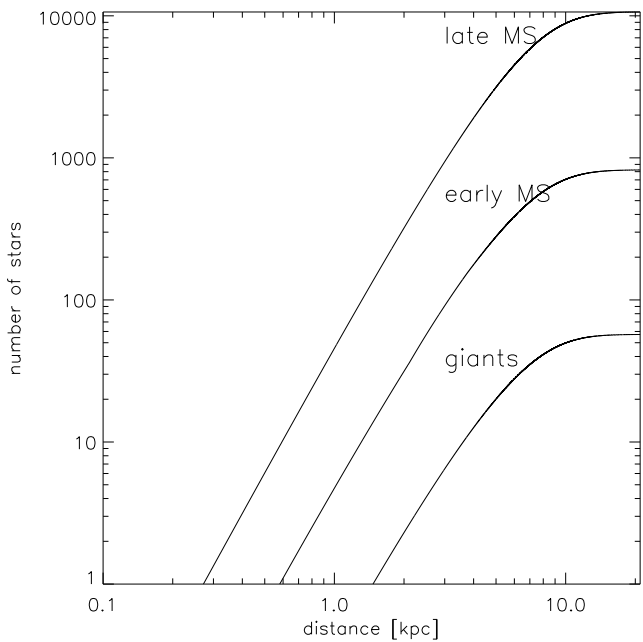}
      \caption{ The accumulated number of stars within 25 square arcminutes as a function of distance as predicted by the
      galaxy model of Wainscoat \& Cohen (\cite{Wainscoat}) for the direction of B\,335.}
         \label{aa11574fig03}
   \end{figure}
%
%______________________________________________________________
%                                                              fig 4

   \begin{figure}
   \centering
   \includegraphics[width=9cm]{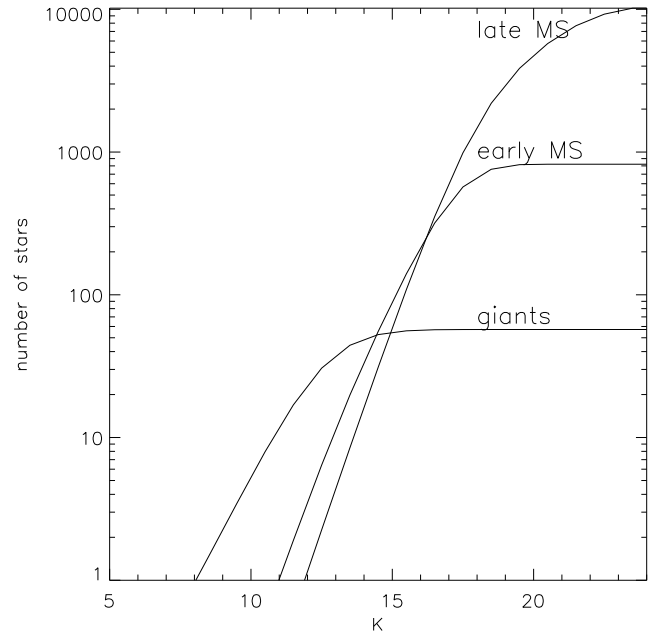}
      \caption{ The K luminosity function for a 25 square arcminutes field as predicted by the
      Galaxy model of Wainscoat \& Cohen (\cite{Wainscoat}) for the direction of B\,335.}
         \label{aa11574fig04}
   \end{figure}

 In a volume limited sample, dwarfs are much more frequent than giants which means that it is very likely that the closest
 stars are dwarf stars. On the other hand, for a given apparent flux limit, the horizon is much
 further away (typically an order of magnitude) for giants than for dwarfs, and giants will be
 over-represented by as much as three orders of magnitude. In order to quantitatively estimate this effect,
 we used the stellar distribution model of the Galaxy as described by Wainscoat and Cohen (\cite{Wainscoat}) for
 the galactic coordinates of B\,335 ({\it l}=44.9, {\it b}=--6.6). In Fig~\ref{aa11574fig03}
 we show the expected number of stars within our field of view as a function of distance. As expected, the
 MS stars dominate. On the other hand, the giants dominate among the brightest stars (Fig~\ref{aa11574fig04}).

 \subsection{How to identify the closest stars?}

\begin{figure}
   \centering
   \includegraphics[width=9.4cm]{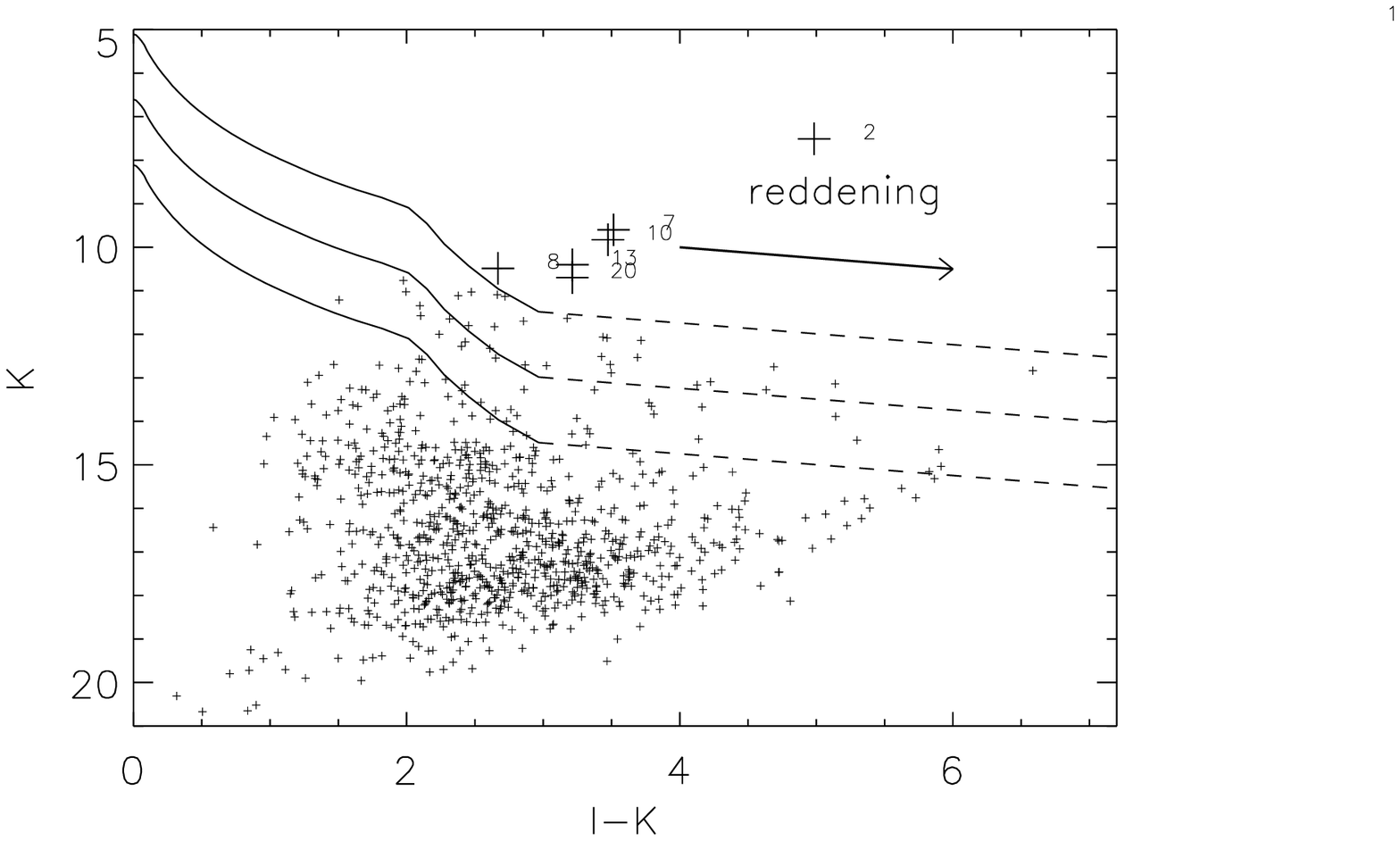}
      \caption{In this infrared HR diagram we have plotted the stars in the B\,335 field.
      The curves represent the main sequence stars at 75, 150 and 300\,pc.
      The reddening vector indicates the effect of increased extinction on a star position.
      The seemingly closest stars are identified by larger symbols and  numbers.}
         \label{aa11574fig05}
   \end{figure}

In  Fig~\ref{aa11574fig05} we display  an infrared HR diagram with I--K as colour (temperature) axis and the
apparent K magnitude as the brightness axis. We also indicate the MS locus for three different distances.
The brightest star in the field is a known variable star (V1142 Aql, equivalent to S 9453 in
the catalogue of Hoffmeister  \cite{Hoffmeister}) and it is probably an AGB star.
Because of its extreme position in the diagram (I--K = 13.6), we left it outside
the plotting range. The positions of the observed stars in the diagram is determined by the
luminosity and the colour as well as the extinction and the distance for each star.
We have also indicated the reddening vector for the K-extinction in the diagram. The slope of the
reddening vector is less than that of the locus for the main-sequence stars in the diagram.
Thus correcting for the extinction will result in still longer distance estimates. In addition, as discussed
above, most of the stars seemingly within 150\,pc may well be giants far behind the cloud.

\begin{figure}
   \centering
   \includegraphics[width=9.4cm]{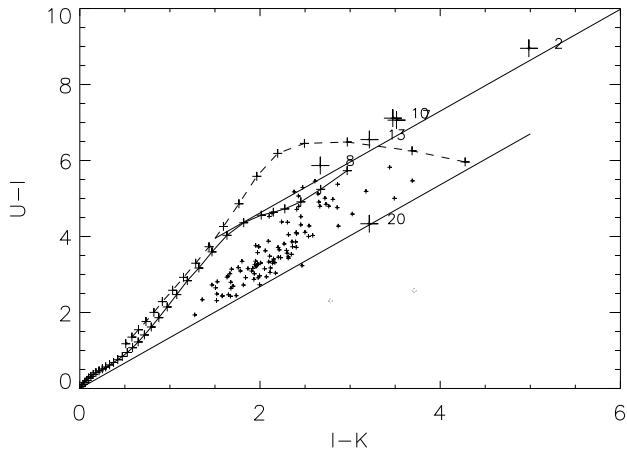}
      \caption{ In this colour-colour diagram we have only included stars with good photometry
            ($\sigma < 0.05$ magnitudes). The seemingly closest stars are marked by larger symbols (cf. Fig~\ref{aa11574fig05}).
            The dashed curve denotes the the expected locus for giants and the full drawn curve the locus for
            MS stars (No. 20 being one of them). The lowest temperature is 3000\,K, and  each 200\,K is indicated.
            The reddening vector is well defined by the locus of early type MS stars.
            All the seemingly closest stars, except No. 20, are most probably giants.}
         \label{aa11574fig06}
   \end{figure}

From Fig~\ref{aa11574fig06} we could conclude that all
the 'nearby' stars appear to be giants, except one which is a heavily obscured early-type star. All
of these are therefore remote, except possibly No.\,2. With the following reasoning we find that
this star also is remote. The reddening line through star No.\,2 crosses the
giant branch at two positions with T$_{eff}\,\approx$\,3400\,K and $\approx$\,4200\,K (which corresponds
to M5III and K4III respectively). If it is a M5III star, the reddening is E(I--K) = 2.0 which
correspond to A$_{K}\approx$\,0.5. The apparent magnitude of star No.\,2 is K=\,7.51 and as
the absolute magnitude of a M5III is M$_{K} \approx$ \,--5, the distance to this star would be $\approx$3\,kpc.

\begin{figure}
   \centering
   \includegraphics[width=9.4cm]{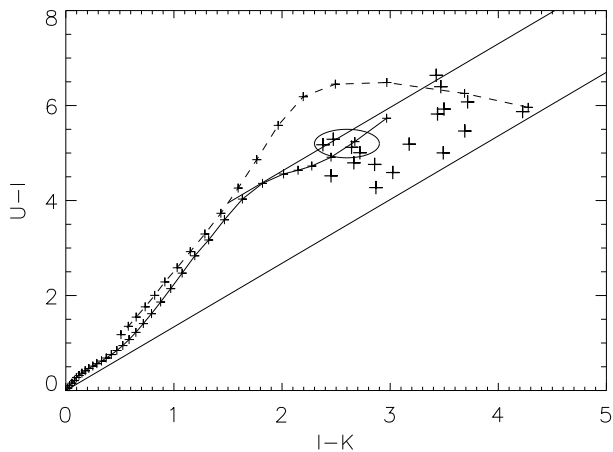}
      \caption{In this colour-colour diagram including the stars within
      the 75 and 150\,pc curves in Fig~\ref{aa11574fig05} there are
      no obvious giants. On the other hand, tracing back along the reddening vector gives
      intersections with both the giant and MS curves, except possibly for the five stars
      closest to the lower reddening line, which appears to be early type MS stars.
      The four encircled stars are potential candidates for foreground M dwarfs.}
      \label{aa11574fig07}
   \end{figure}

\begin{figure}
   \centering
   \includegraphics[width=9.4cm]{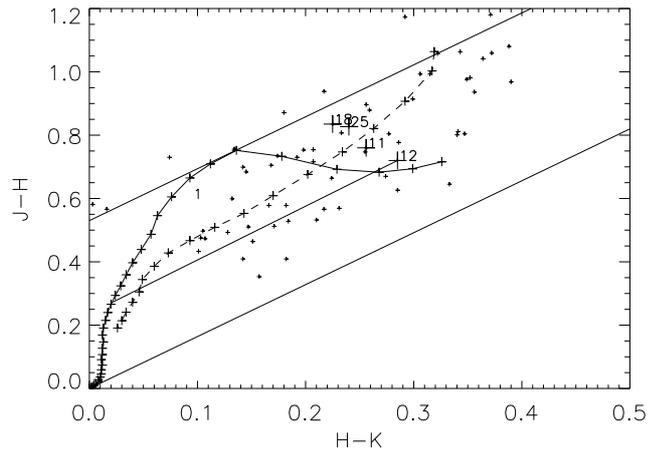}
      \caption{In this J - H colour vs H - K colour diagram the four stars
      identified in Fig~\ref{aa11574fig07} as possible foreground stars
      are marked with larger symbols. We argue that one of them, No. 12, is
      indeed a foreground M dwarf.}
         \label{aa11574fig08}
   \end{figure}

We next turned to the 23 stars seemingly at a distance between 75 and 150\,pc and display them in a
U--I versus I--K diagram (Fig~\ref{aa11574fig07}). Except for the four
stars closest to the lower reddening line (which probably are early-type MS stars),
it is not certain whether to designate the stars as dwarfs or  giants
because the reddening vectors intersect both the locus of the giants and that of the dwarfs.
Although the general impression is that the stars are heavily obscured (and therefore behind the cloud), we
could not exclude that a star close to the cool part of the MS locus actually is a foreground star. If this
were the case, its distance would represent a lower limit of the distance to the cloud. To further constrain
the properties of these stars we obviously needed more information and as we, so far, only had considered one
colour--colour diagram out of several possible it is natural to look for some other combination. It is well known that the
continuum opacity minimum at 1.6\,$\mu$m is pressure dependent, which is reflected in the J--H versus
H--K diagram (Fig~\ref{aa11574fig08}). We marked the position of the four possible foreground stars in this diagram, and indeed,
one of them (No. 12) can either be a late M dwarf or (if behind the cloud) a star with
T$_{eff}\,\approx$\,\,6000\,K. However, such a high temperature can be {\it excluded} from the position of this star in
the U--I versus I--K diagram, and this star is therefore very likely a cool  foreground  star.

\subsection{A more efficient approach}

In principle it would be possible to further constrain the properties of the stars by
inspection of other colour combinations, but it is not a very practical way to proceed. In addition, we have so far
assumed that the reddening law is independent of the light path through the cloud, be it close
to the center or in the outskirts. This is not likely to be the real case but, as we will
show in a forthcoming paper, the scheme introduced by Cardelli et al.(\cite{Cardelli})
where the extinction is defined by a single parameter $R_V$, adequately represents the extinction.

\begin{figure}
   \centering
   \includegraphics[width=9.0cm]{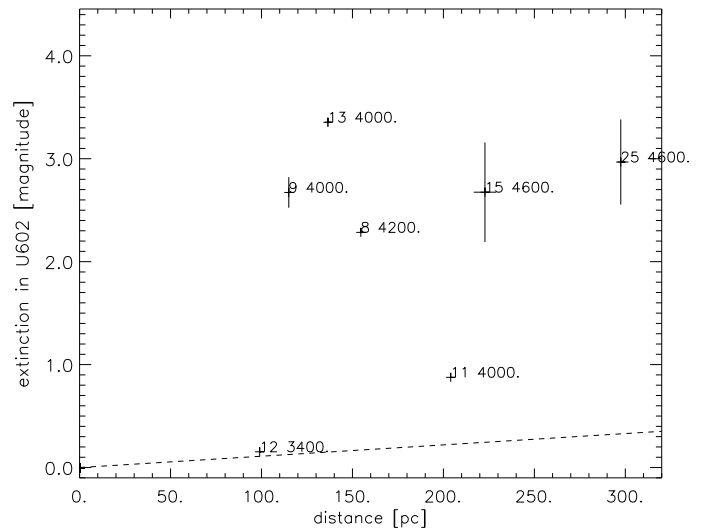}
      \caption{ This diagram shows the distances and the extinctions
      to the closest stars in the B\,335 field. Only one star is found
      in front of the cloud. The error bars only reflect the $\chi$-square
      fitting errors, and the indicated distance to B\,335 is 90--120\,pc.
      The dashed line schematically represents the  extinction due to the diffuse ISM.}
      \label{aa11574fig09}
   \end{figure}

A more general and efficient approach would be to start by searching for stars that are likely to be MS stars.
With this step we sought the best $\chi$-square fit of the SED for each star among the  61 model SEDs,
optimising for the R$_V$ parameter and the extinction (A$_U$). This optimisation is fast and simple, but one
may wonder whether it is reasonable to let the R$_V$ parameter be a free parameter for each star. Actually,
when correlating the resulting $R_V$ parameters to the positions of the stars, we saw a tendency in the sense that
the $R_{V}$-values are higher in the eastern part than in the western. This will be discussed further
in a forthcoming paper, but more significant for the present analysis is that we noted a high scatter regarding
the $R_{V}$-values for adjacent stars. This is not likely to reflect the true conditions, and therefore we
introduced a second step. In this step we selected probable nearby MS stars found in the first step. For each of
them we repeat the optimisation, but this time together with a few adjacent stars requesting that
this grouping experiences a cloud extinction characterised by the {\it same} R$_{V}$-value. We allowed,
however, for an individual total extinction, A$_U$. Finally, in order to get a measure on the stability of
the solutions we repeated the calculations a few times for different combinations of adjacent stars. The results are shown in
Fig~\ref{aa11574fig09}. We first noted that star No.\,12 also in this analysis is found to be a
foreground star, and its photometric distance is 90\,pc. The two closest stars behind the cloud is No.\,9 (120 \,pc) and
No.\,13 (140\,pc). The distance to B\,335 is therefore 90--120\,pc, neglecting possible systematic errors.

\subsection{Can we trust the photometric classification?}
We classified the collected spectra by comparison to templates from the ESO/UVES survey (\cite{UVES}).
The temperature, T$_{eff}$, is then taken from the compilation in Allen's Astrophysical Quantities (\cite{Allen}).
The photometric classification gives T$_{eff}$ and the preferred status
giant-or-dwarf, and for convenience we used the same calibration to indicate the spectral class. The result
is shown in Table~\ref{spectral_classes}. The agreement is in most cases good.\\
It is, however, worrying that the only star in
the list that may be of interest for the distance estimate, No 13, differs considerably.
The TiO and CaH bands are pronounced so it is definitely a M star and the NaI lines indicate that it is a dwarf star.
However, it cannot be completely excluded that it may be a giant and
that the strength of the sodium doublet is due to interstellar absorption components.
In either case, this star is not relevant for our distance determination, as it not the closest star behind the cloud.
%%%which must be modified to {\bf 90 -- 140\,pc. }
\begin{table}
\begin{minipage}[t]{\columnwidth}
\caption{Spectral classification}
\label{spectral_classes}
\renewcommand{\footnoterule}{}  % to avoid a line before footnotes
\begin{tabular}{rllll}

\hline \hline

\textbf{star}&\textbf{spectrometric}&&\textbf{photometric}\\

&\textbf{Teff}&\textbf{MK}&\textbf{Teff}&\textbf{MK}\\

2&3850&K7III&3900&K6III\\

7&3850&K7III&4300&K3III\\

10&3850&K7III&4100&K4III\\

13&3300&M3V&4000&K7V\\

20&5940&G0V&6400&F6V\\

84&4650&K0III&4800&K2V\\

112&6250&F8V&5600&G8V\\

134&6200&F8V&6400&F6III\\

137&5800&G0V&5900&G0III\\

214&4050&K5III&4900&G8III\\

\hline\\

\end{tabular}

\end{minipage}
\end{table}

%\begin{tabular}{rllllll}    %%%{rrrrrrr}
%\hline
%star&M-K&$T_{eff}$&extinct&recta&dec&distance\\
%num&cat&$^{o}K$&$A_{U}$&J2000&J2000&$[pc]$\\
%29&K6V&4200&3.5&294.2218&7.5564&$150\ \pm \ 5$\\
%8&K8V&4000&2.0&294.2971&7.5640&$152\ \pm \ 5$\\
%(3&K0V&5200&2.2&294.2711&7.5668&$161\ \pm 10$)\\
%\hline\\
%\\
%\end{tabular}
%%\end{minipage}
%%\end{table}
%______________________________________________________________
%                                                              fig 5

%

%______________________________________________________________
%                                                              fig 6

%

%______________________________________________________________
%                                                              fig 7

%
%______________________________________________________________
%                                                              fig 8
%
%______________________________________________________________
%                                                             fig 9
%
%_____________________________________________________________
%                                                              fig 10
   \begin{figure}
   \centering
   \includegraphics[width=9.0cm]{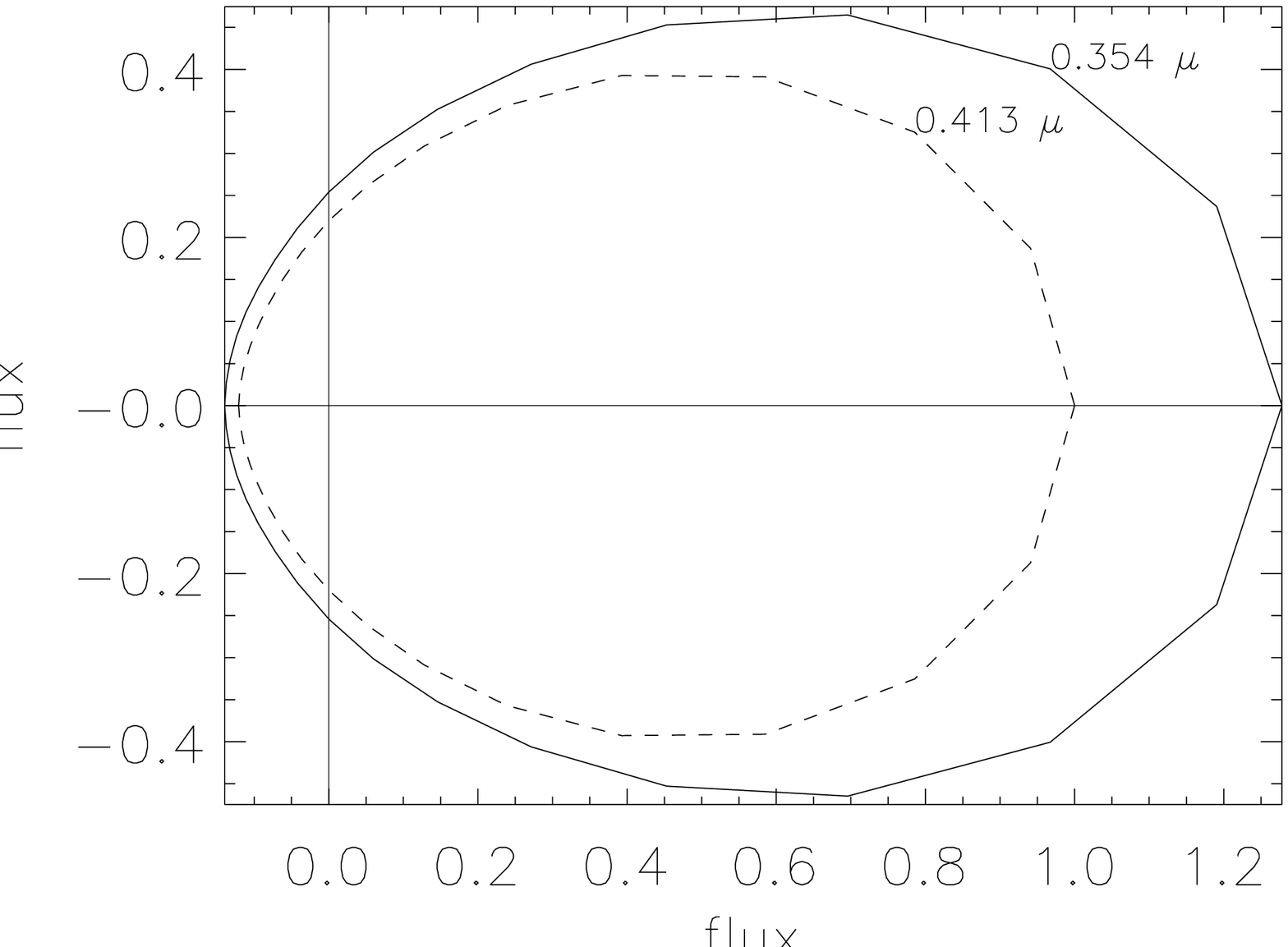}
      \caption{ The phase diagram for the MRN grain size distribution model.
      It is normalised for the B filter at a scattering angle of zero degrees.
      The diagram shows that the forward scattering is much more efficient
      than the back scattering and also, that this is more pronounced in the U than in B band.}
      \label{aa11574fig10}
   \end{figure}
%
%_____________________________________________________________

\subsection{The bright rim of B\,335}

Dark globules often show bright rims due to scattered light originating from the surrounding
illumination by the interstellar radiation field. One may wonder why the core of the globule
looks dark as it is also illuminated by the interstellar radiation field.
The reason is the angular dependence of the scattering, being more efficient in the forward direction.
To illustrate this point we calculated the phase diagram for the grain size distribution model (MRN)
given by Mathis et al (\cite{MRN}).

In Fig~\ref{aa11574fig10} we show the scattering efficiency as a function of angle, and we include both the wavelength for the
U and the B band. The forward scattering is $\approx$ factor of 4 larger than the back scattering, explaining why the rim of a dark
globule in general is bright. It must also be realized that it is virtually impossible to
determine the sky emission accurately enough to measure the light level at the center of a dark globule.
In the case of B\,335, we do observe a bright rim and we note a pronounced enhancement
in the south-west indicating that there is a local
enhancement due to a nearby star. We first considered the case that this star illuminates the globule. We next
investigated the possibility that the light enhancement is due to the  point spread function (PSF) of a bright
star located outside our field. The most probable star (in both cases) is HD\,184982 which is located 1.5 arcminutes
to the south-west of the image. This star has an entry in the Hipparcos Catalogue and its distance is 140--200\,pc.

\begin{figure}
    \centering
    \includegraphics[width=9.5cm]{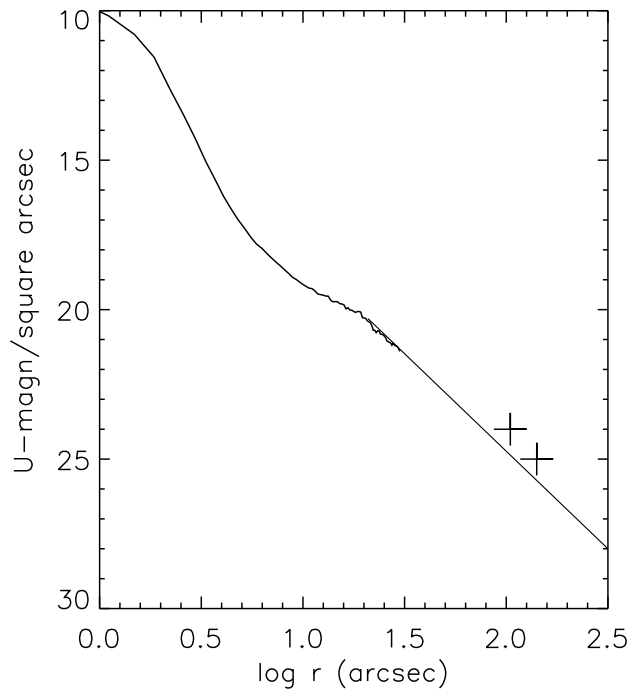}
    \caption{The point-spread-function of HD\,184982. The extrapolated PSF wing falls slightly
    below the measured surface brightness at the south-western rim of B\,335 (crosses).}
    \label{aa11574fig12}
\end{figure}

As the center of the globule is dark,
this star -- if physically associated --  must be located behind the cloud, but how far behind? We first note that the
surface brightness close to the south-western edge of the image, U = 24 magnitudes\,arcsec$^{-2}$, corresponds to
0.4 $\mu$Jy\,arccsec $^{-2}$, which is comparable to the interstellar radiation field, 1.0\,$\mu$Jy\,arcsec$^{-2}$.
We must, however, keep in mind that the rim of the globule only is illuminated from one side
and that the albedo of the grains is less than unity. Therefore, additional illumination is required to explain the bright
south-western rim of the globule. The A2V star has an absolute magnitude M(U)$\approx$1.4 which means
that it gives the same (one-side) illumination as the interstellar radiation field if it
were located 0.4 pc from the cloud. It is likely that the star is located a bit closer to the cloud in order to
give rise to the observed illumination, but for our purpose it suffices to conclude that the distance to the B\,335 globule
is essentially the same as that to HD\,184982 (provided that it really is physically associated).
We now turned to the  possibility that we just measured the wing of the point spread function (PSF)
of this star. By co-adding 23 images of the calibration star, we are able to define the PSF to a radius
of 40 arcseconds. A power-law extrapolation showed that the observed rim has a surface brightness of a factor 2 higher
than expected from the PSF (Fig~\ref{aa11574fig12}). The index of the power law extrapolation, --2.6,
makes sense, but in view of how far (from a radius of 40 to 100 arcseconds) we had extrapolated the measured
PFS, we could not claim that the difference between the measured surface brightness at the SW rim of B\,335 is significantly
higher than the PSF wing of  HD\,184982.
Therefore this explanation is (unfortunately) more likely than the physical association.
If the photometric distance to the globule of 90--120\,pc is accepted the star HD\,184982 (with its
Str\"omgren photometric distance of 127\,pc) is still one of the stars close to the cloud.
%%If the interpretation of the enhanced brightness of the south-western rim of the
%%globule not being due to the reflected light from  HD\,184982, then we can conclude that the photometric
%%distance as determined by the closest stars in the field, 90--120\,pc, agrees well with that determined
%%by means of  Str\"omgren photometry (127\,pc) and just touches the closer limit, 140\, pc, of
%%the Hipparcos distance.

%______________________________________________________________

   \begin{figure}
   \centering
   \includegraphics[width=9.0cm]{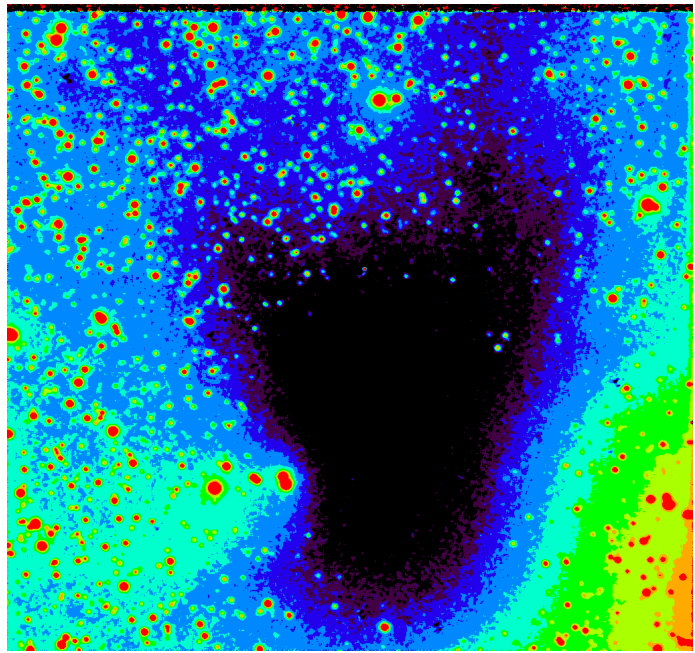}
      \caption{ The U image of B\,335 shows a rim of scattered light as is
      often the case for dark globules. The pronounced enhancement in the
      SW corner is proposed to be due to a bright star 1\farcm5 outside
      the field. The lowest contour level corresponds to U\,=\,27 arcsec$^2$
      and the step size is 0.5 magn }
         \label{aa11574fig11}
   \end{figure}
%
%______________________________________________________________

\section{Discussion}

\subsection{How to apply the method to a large sample of globules?}
If the Galactic stellar distribution model by Wainscoat \& Cohen (\cite{Wainscoat}) applies,
then we were very lucky to find four stars in our field being closer than 200\,pc.
The model may underestimate the low-mass population of stars, and if so, our field of view
($\approx$36 square arcminutes) may suffice. Still, in planning for a survey of dark globules
aiming at distance determinations, it would be better to observe larger fields.
However, in most cases, the globules as seen in the optical are small and it is often
difficult to define how far their halos extend. Even though the UV imaging is more efficient
in revealing the extent of the globule (including its bright rim), a separate CO survey designed to define the sizes
would  be very useful. In our pilot study we spent a very long time on the UV imaging. This
was justified by the expectation that the closest stars would likely be M dwarfs and therefore faint in the UV and,
in addition, the extinction of the globule is higher at short wavelengths. The faintest of the nine closest
stars, No 35, has a UV brightness of U = 20.00 ($\sigma$=0.014), and as a photometric accuracy
of $\sigma$=0.04 is estimated to be good enough, 1\,h exposure time should have sufficed (instead of the
8\,h spent). Also for the other photometric bands in the optical, the exposure times could have been
much shorter. The faintest star in the K$_s$ band among these nine closest star, No 25, is bright,
K$_s$  = 11.75 ($\sigma$=0.002), and consequently the 2MASS survey provides the required accuracy. This all means that typically
2--3 hours per globule should suffice for a distance survey using a wide-field camera on a medium sized telescope.

\subsection{Is the method reliable?}

It is clear from this test case that a larger field than our 36 square
arc-minutes should be used in order to include more stars at the relevant distances. The photometric classification is
in most cases adequate, but we noted exceptions. One reason for exceptions is probably significantly lower metal abundance.
It is known (Buser \& Kurucz, \cite{Buser} ) that colour indices including the U band do depend on the metal abundance. In order to
quantify this effect we have used a set of model atmospheres with $[$Fe/H$]$\,=\,--1, and the
difference compared to solar abundance in the U--I colour index is 0.3 magnitudes for late M stars.
This is far more than the photometric accuracy. It is likely that the mismatch for
star No 13, discussed above, is caused by a low metal abundance. Again, a larger field would include more stars at a
relevant distance and make it easier to exclude stars with very different metal abundance compared to the majority of the stars. In
any case, follow-up spectroscopic observations of the stars identified by the photometry as being located closely in front and
behind the cloud would be useful for verifying (refining) the spectral classification.

One obvious source of uncertainty is
non-resolved binaries. Thus, if the stars have approximately the same spectral class, the inferred distance
would be too small by a factor $\sqrt2$. Otherwise the error will of course be less, and there is also
a chance that our SED-fitting procedure may flag such cases (simply by a worse fit),
but we have not looked closer at this aspect in our pilot study. But, again, this
complication is less of a problem when a larger field is applicable improving the statistics.

\subsection{Is the method computationally practical?}

The first step of our method, i.e. finding the best SED fit with the distance and the R$_{V}$-value as parameters,
is fast and simple. The reason why we found a need for a second step is the resulting large
scatter of the R$_{V}$-value between adjacent stars, which is not likely to be real. To cope with this
problem, we added the second step where we carried out the SED fitting for adjacent stars {\it requesting}
that they share the same R$_{V}$-value. The computational time for this step is proportional to the
N$th$ power, where N is the number of stars in each group. In order to get a measure on the
stability of a fit, we also ran the optimisation for different combinations of adjacent stars.
In practice we found that N = 3 is sufficient and that the computational time can be kept low.

\section{Conclusions}

   \begin{enumerate}
      \item The fundamental limitation when it comes to determining the distance to a small dark cloud is the space density of stars and as the cool MS stars are the most abundant, any method should
      be able to sort out these stars and reliably separate them from cool giant stars.
      \item It turns out that multi-colour imaging in combination with synthetic colours can provide reliable spectral classification and extinction, and we find that the
      U band is particularly important, both for the spectral classification and for tracing the extent of the globule.
      \item We have tested a new SED-fitting method which allows for different R$_{V}$-values
      across the field. For the test case, B\,335, we get a distance of 90--120\,pc.
      \item Our deep U image of B\,335 shows a bright halo due to scattered star light, and the south-western rim is much brighter. We find that a possible explanation is that the globule is illuminated by a star, HD\,184982, located 1.5 arcminutes  to the south-west of our field and a few tenths of a parcec behind the cloud. The distance to this star
      is 140--200\,pc according to the Hipparcos Catalogue. However, the surface brightness can also be explained as
      due to the PSF wing of the same star, and we judge that this is in fact a more likely explanation.
   \end{enumerate}

\begin{acknowledgements}
\vspace{0.5 cm}
Peter H Hauschildt, Universit\"{a}t Hamburg, has kindly provided us with the stellar atmosphere models.\\
Magnus G{\aa}lfalk, Stockholm University, has contributed the deep NOTCam K$_{s}$ image of B\,335.\\
This publication makes use of data products from the Two Micron All Sky Survey, which is a joint project of
the University of Massachusetts and the Infrared Processing and Analysis Center/California Institute of Technology,
funded by the National Aeronautics and Space Administration and the National Science Foundation.\\
\end{acknowledgements}

\end{document}